\documentstyle[12pt]{article}
\begin{document}
\def\bfm#1{{\mbox{\boldmath $#1$}}}
\def\bfms#1{{\bfm{\scriptstyle #1}}}
\def\mezzo{\frac{1}{2}}
\def\intq#1{\int \frac{d^4#1}{(2\pi)^4}}
\def\intt#1{\int \frac{d^3#1}{(2\pi)^3}}
\def\Tr{{\rm Tr}}
\def\k{{k_F}}
\def\square{\hbox{\framebox{~}}}
\def\ve{\varepsilon_l({\bf k},{\bf p})}
\def\nd{<{\bf p},A|{\bf p},A>}
\def\phil{\phi_l({\bf k},{\bf p})}
\def\chil{\chi_l({\bf k},{\bf p})}
\def\G{G_{\bfms{p}}(k)}
\def\GG{{\cal G}_{\bfms{p}}}
\def\cl{C_{l}({\bf k},{\bf p})}
\def\dki{\frac{d^3k_i}{(2\pi)^3}}
\def\ppsi#1{\Psi(#1)}
\def\ppsid#1{\Psi^\dagger_(#1)}
\author{R. Cenni$^\dagger$, A. Molinari$^\ddagger$ and G. Vagradov$^*$
\\~\\~\\
$^\dagger$Dipartimento di Fisica dell'Universit\`a di Genova \\
Istituto Nazionale Fisica Nucleare -- Sezione di Genova \\~\\
$^\ddagger$Ministero degli Affari Esteri \\
Consolato Generale d'Italia -- Boston -- U.S.A.\\~ \\
$^*$
Institute for Nuclear Research\\
117312 Moscow Russia}
\title{On the Relativistic Description of the Nucleus}
\maketitle\hfill

\begin{abstract}
We discuss a relativistic theory of the atomic nuclei in the
framework of the hamiltonian formalism and of the mesonic model of the
nucleus.
Attention is paid to the translational invariance of the theory.
Our approach is centered on the concept of spectral amplitude, a
function in the Dirac spinor space. We derive a Lorentz covariant
equation for the latter,  which requires as an input the baryon
self-energy.
For this we either postulate the most general Lorentz-Poincar\'e
invariant expression or perform a calculation via a Bethe-Salpeter
equation starting from a nucleon-nucleus interaction. We discuss the
features of the nuclear spectrum obtained in the first instance.
Finally the general constraints the self-energy should satisfy because
of analyticity and Poincar\'e covariance are discussed.
\end{abstract}

\newpage

\section{Introduction\label{sec1}}

The purpose of this paper, which extends and deepens a previous
one\cite{MoVa-89}, is to outline a theory
of finite nuclei cast in a
Lorentz and translationally invariant form.

Extensive work
has been carried out in the past to describe a  nucleus in
a relativistic regime (see for instance \cite{SeWa-86,CeSh-86-B} for
comprehensive reviews).
This has been done essentially in the framework of the Dirac's theory,
either phenomenologically\cite{Wa-74,ClHaMe-82,Cl-86}, i.e.
by suitably parametrizing the self-energy felt by a relativistic nucleon
when interacting with a nucleus, or microscopically, i.e. by evaluating the
self-energy starting from a mesonic theory, like
the Bonn potential \cite{Ho-81,MaHoEl-87}).

However, the Dirac's theory, as it is usually applied,
breaks translational invariance.
This appears to be serious at high energies as
already indicated by nonrelativistic calculations, which point to the
growing importance of the center-of-mass component in the nuclear wave
function when the momentum transfer becomes large\cite{TaBa-58}.

A notable extension of the Dirac's theory is represented by the
Quantum Hadrodynamics (QHD) of Walecka\cite{BiSe-84,HoSe-87},
a quantum field theory model based on a lagrangian which
describes nucleons interacting via the exchange of isoscalar scalar
($\sigma$) and vector ($\omega$) meson fields.

Notably, QHD has been remarkably successful in dealing both with nuclear
matter and finite nuclei\cite{HoSe-81} in the so-called mean field
approximation (MFA), where one replaces the meson field operators by
their ground state expectation values ignoring furthermore the
negative energy states. However when one goes beyond a MFA to include
correlations among nucleons (via, e.g., a Bethe-Salpeter-type
equation or truly quantum field theory effects, as those stemming
from vacuum polarization) then QHD faces severe problems. Indeed,
although great care is paid in its framework to the problem of
renormalization, the loop expansion appears to be, to say the least, quite
poorly convergent. For example the two loop contribution to the binding energy
turns out to be very large at the nuclear matter density\cite{PeWa-88} because
of the unsatisfactory ultraviolet behaviour of the theory, a well known feature
common to all the non asymptotically free model. One can of course deal
with these difficulties by introducing {\em ad hoc} form factors to cut off
the large momenta entering the diagrams needed to express the
physical observables of concern, but in so doing one spoils QHD of its
character of a truly hadronic field theory.

Our approach differs from QHD in several respects. First it is centered
on the concept of single particle
orbital. In fact, notwithstanding the difficulty of defining a
wave function, in particular a single
particle one, in a relativistic framework,
yet it is still  advantageous to deal with the nucleus in terms of a
relativistic extension of the
quasiparticles, the building blocks of atomic nuclei,
as introduced by Landau and then by
Migdal\cite{Mi-67-B}.

To achieve this goal it is crucial
to introduce the so called spectral amplitude $\phi_l$, to be
later defined, a suitable tool for expressing the nuclear binding
energy, the nucleon momentum distribution and therefore
the cross sections for a variety of nuclear
reactions (for example the exclusive $(e,e'p)$ one\cite{PaSa-82})
in a Poincar\'e invariant form.

A second difference from QHD arises from the recognition that
the connections between $\phi_l$ and the observables above
referred to are more naturally established
in a {\em hamiltonian} framework where they can be cast in the
form of {\em exact} relations. Of course
the problem of the infinities arising from the presence of the Dirac's sea
exists and we deal with it empirically
by means of {\em ad hoc} subtractions.

Moreover the hamiltonian scheme has the advantage of allowing
us to fulfill more easily (at least formally)
Poincar\'e invariance when treating finite nuclei. As a notable
consequence the Dyson's equation for the fermion propagator, from which
an equation for $\phi_l$ is derived, exhibits
an algebraic, rather than an integral, structure even for
finite nuclei. Obviously here the difficulty is met of properly defining and
calculating the self-energy of a baryon or of a meson (see sect. \ref{sec4}).

In this connection we adopt a pragmatic view, namely we write the most
general Poincar\'e invariant self-energy in terms of functions meant to be
fixed
by the phenomenology\cite{Wa-74,Mi-75,ClHaMe-82,Cl-86} and we explore the
consequences it entails on the spectrum of the system.

We also investigate the alternative option of {\em calculating}
the self-energy: here our approach is based on a Bethe-Salpeter-type
equation\cite{CeSh-86-B},
with a phenomenological nucleon-nucleus interaction as an input.
We thus account, respecting covariance, for the repeated interactions
a particle undergoes in its way-out from the nucleus (final state interaction).
It is worth noticing that this approach
has the merit of transparently displaying for a convenient choice
of the interaction and  for an infinite homogeneous system how
our hamiltonian theory
reduces to QHD at the mean field level.

This paper is organized as follows: in sect. \ref{sec2} we outline our
hamiltonian framework, introduce the spectral amplitude and establish
the connection of
the latter with various physical observables. In sect. \ref{sec3} we
discuss a simple example to illustrate how mesonic degrees of freedom
can be accounted for by means of a canonical transformation. In sect.
\ref{sec4} we study the algebraic Poincar\'e-covariant Dyson's
equation for the fermion propagator in a finite system. The
Lehmann representation for the propagator is also analyzed.
We study then the general structure of the
self-energy, its link with the final state interaction via
a Bethe-Salpeter equation and its analytic properties. Finally
we shortly analyze the resulting energy spectrum of the nuclear system.

\section{The formalism\label{sec2}}

Let us consider a system of nucleons interacting via the exchange of mesons.
For simplicity we confine ourselves in this section to
consider the exchange of a scalar meson. Euristically the latter simulates
the exchange of a pair of pions, a microscopic process viewed as the main
source of the nuclear binding.
Clearly our model should then be
generalized to more realistic situations encompassing
several different mesonic fields to account, e.g., for the short range
repulsion
among nucleons.

Its hamiltonian reads
\begin{equation}
{H}={H}^0_N+{H}^0_\sigma+{H}'\label{v1}
\end{equation}
where
\begin{eqnarray}
{H}^0_N&=&\int d^3y \,\overline{\psi}({\bf y})(-i\bfm{\gamma}\cdot
\nabla+m)\psi({\bf y})\label{v2a}\\
{H}^0_\sigma&=&\mezzo\int d^3y\,\left[\dot\sigma^2({\bf
y})
+(\nabla\sigma({\bf y}))^2+m^2_\sigma\sigma^2({\bf y})\right]\label{v2b}\\
{H}'&=&g\int d^3y\,\overline{\psi}({\bf y})\sigma({\bf y})\psi({\bf y})
\label{v2c}\;,
\end{eqnarray}
$\psi$ and $\sigma$ being the fermion and meson field.

A flaw of (\ref{v1}) stems from
the absence of a self-interacting term $\lambda\sigma^4$, needed
to bound from below the spectrum of (\ref{v1}). However since,
as we shall see in the following, the stability of the
$\sigma$ field is here ensured by the presence of the nuclear medium,
the renormalized value of $\lambda$ can be assumed to be positive and
small: hence the  $\lambda\sigma^4$ term is not expected to significantly
affect the predictions of the theory and can thus be safely omitted.

A further shortcoming of (\ref{v1}) is connected with the
difficulty of defining a vacuum for a bound, finite, relativistic Fermi system.
Indeed the standard expansion for a fermion field,
namely
\begin{equation}
\psi({\bf y})=
\sum_{s}\intt{k}{m\over\epsilon_k}
\left[u_{ks}{a}_{ks}e^{i\bfm{\scriptstyle k}\cdot\bfm{\scriptstyle y}}
+v_{ks}{b}^\dagger_{ks}e^{-i\bfm{\scriptstyle k}\cdot\bfm{\scriptstyle y}}
\right]\label{v4}
\end{equation}
(${a}_{ks}$ and ${b}_{ks}$ being the free nucleon and
antinucleon annihilation operators and $\epsilon_k=\sqrt{{\bf
k}^2+m^2}$), is not appropriate, being incompatible with the
spatial confinement of the atomic nucleus.

However, since we do not pretend to obtain here a solution of the problem
of confined states in QFT (in fact we rather search for relations
between physical observables and the spectral amplitude)
we shall still quantize
the fermion field according to (\ref{v4}), but without introducing
normal products at all. We shall instead resort to the introduction of
compensating terms, in the form of vacuum subtractions, in order to avoid
the occurrence of infinities.

On the other hand, as far as the mesons are concerned, we assume their
coupling with the fermion field to be so weak to prevent the
occurrence of bound meson states. Therefore the expansion
\begin{equation}
\sigma({\bf y})=\intt{k}\frac{1}{2\omega_k}
\left[{c}_{k}e^{i\bfm{\scriptstyle k}\cdot\bfm{\scriptstyle y}}
+{c}^\dagger_{k}e^{-i\bfm{\scriptstyle k}\cdot\bfm{\scriptstyle y}}
\right]\label{v5}
\end{equation}
(${c}_k$ and ${c}^\dagger_k$ being the free meson annihilation
and creation operators and $\omega_k=\sqrt{{\bf k}^2+m^2_\sigma}$) is
likely to be still warranted.

In line with the above  considerations we shall rewrite
equations (\ref{v2b}) and (\ref{v2c}) as follows:
\begin{eqnarray}
{H}^0_\sigma&=&\mezzo\int d^3y:\left[\dot\sigma^2({\bf y})
+(\nabla\sigma({\bf y}))^2+m^2_\sigma\sigma^2({\bf y})\right]:
\nonumber\\
&=&\int\frac{d^3k}{2\omega_k(2\pi)^3}\,\omega_k c^\dagger_kc_k\label{v2b'}\\
{H}'&=&\int\frac{d^3k}{2\omega_k(2\pi)^3}\left(c^\dagger_k\rho_k+
\rho^\dagger_kc_k\right)\label{v2c'}
\end{eqnarray}
where
\begin{equation}
\rho_k=g\int d^3y\,\overline{\psi}({\bf y})\psi({\bf y})e^{-i
\bfm{\scriptstyle k}\cdot\bfm{\scriptstyle y}}\label{v8}\;.
\end{equation}
Our choice of the normalization
of the fields is such that the commutator
\begin{equation}
[c_k,c^\dagger_{k'}]=(2\pi)^32\omega_k\delta({\bf k}-{\bf k}')\;,
\label{v3n}
\end{equation}
holds for the boson
and the anticommutator
\begin{equation}
\left\{a^\dagger_{ks},a_{k's'}\right\}=(2\pi)^3\frac{\epsilon_k}{m}
\delta_{ss'}\delta({\bf k}-{\bf k}')\label{v6n}
\end{equation}
for the fermion fields.

Let us now introduce the nuclear states $|{\bf p},(A)_m>$, identified by
the total 3-momentum $\bf p$, baryonic number $A$ and by a set of
internal quantum numbers denoted by $m$. With the notation
$|{\bf p},A>$ we refer to the internal ground state. Since
the relativistic normalization of a state is
statistic-dependent, we have to choose the parity of $A$. Let us assume $A$
to be even.
Then the normalization reads
\begin{equation}
<{\bf p},A|{\bf p}',A>=(2\pi)^32p_0\delta({\bf p}-{\bf p}')\label{v9c}\;.
\end{equation}
Of course for the ground state
\begin{equation}
{P}_\mu|{\bf p},A>=p_\mu|{\bf p},A>\label{v9a}
\end{equation}
and
\begin{equation}
p_0=\frac{<{\bf p},A|{H}|{\bf p},A>}{<{\bf p},A|{\bf p},A>}
-<0|{H}|0>=\sqrt{M_A^2+{\bf p}^2}\label{v9b}\;.
\end{equation}
Clearly $<{\bf p},A|{H}|{\bf p},A>$ diverges owing to the
contribution to the energy arising from the Dirac sea, hence the vacuum
subtraction in (\ref{v9b}).

In parallel to (\ref{v9a}) for a $(A-1)$ system, left in a quantum state
identified by the index $l$, we have
\begin{equation}
{P}_\mu|{\bf p},(A-1)_l>=p^l_\mu|{\bf p},(A-1)_l>\label{v10a}\;,
\end{equation}
where
\begin{equation}
p_\mu^l\equiv (p_0^l,{\bf p})\label{v10a1}
\end{equation}
and
\begin{equation}
p_0^l\equiv\epsilon^{A-1}_l({\bf p})=\sqrt{(M_l^{A-1})^2+{\bf p}^2}\;,
\label{v10b}
\end{equation}
but now the fermionic normalization
\begin{equation}
<{\bf p},(A-1)_l|{\bf p}',(A-1)_{l'}>=(2\pi)^3\frac{\epsilon^{A-1}_l({\bf
p})}{M_l^{A-1}}\delta_{ll'}\delta({\bf p}-{\bf p}')
\end{equation}
holds.

Let us now introduce in the Dirac spinor space the function
\begin{equation}
\phi_{l\bfm{\scriptstyle k}\bfm{\scriptstyle p}}(y)
=\cl<{\bf p}-{\bf k},(A-1)_l|
\psi(y)|{\bf p},A>\label{v11a}
\end{equation}
with
\begin{equation}
\cl=\sqrt{\frac{M_l^{A-1}}{
2p_0\epsilon^{A-1}_l({\bf p}-{\bf k})}}\label{v11b}\;,
\end{equation}
the field $\psi(y)$ being now in Heisenberg picture.
The above  is the key quantity of our formalism, in terms of which
we shall express the ground state energy and, in a  parallel research,
the structure functions
of the nucleus. In sect. \ref{sec4} we shall derive
the  equation obeyed by $\phi_l$.

The amplitude $\phi_{l\bfm{\scriptstyle k}\bfm{\scriptstyle p}}(y)$
is the tool we need to generalize the quasiparticle concept to a relativistic
regime:  in fact it can be shown that
in the nonrelativistic limit and in the single-particle
approximation $\phi_l$ exactly reduces to a one particle wave function.
Furthermore the coefficient $\cl$ is such to account
for the difference between the relativistic and nonrelativistic
normalization of the states.

Of course the spatial dependence of $\phi_{l\bfm{\scriptstyle k}
\bfm{\scriptstyle p}}(y)$ is fixed by translation invariance. Indeed, from
\begin{equation}
\psi(y)=e^{i{P}\cdot y}\psi(0)e^{-i{P}\cdot y}\label{v5n}
\end{equation}
one gets
\begin{equation}
\phi_{l\bfm{\scriptstyle k}\bfm{\scriptstyle p}}(y)=\phi_l({\bf k},{\bf p})
e^{-ik^l\cdot y}\label{v12a1}
\end{equation}
where
\begin{equation}
\phi_l({\bf k},{\bf p})=\cl
<{\bf p}-{\bf k},(A-1)_l|\psi(0)|{\bf p},A>\label{v12a2}
\end{equation}
and
\begin{eqnarray}
k^l&\equiv&(\ve,{\bf k})\label{v12b1}\\
\ve&=&p_0-\epsilon^{A-1}_l({\bf p}-{\bf k})\label{v12b2}\;.
\end{eqnarray}
Owing to the definition (\ref{v11a}) $\phil$ transforms under the
Lorentz group as a Dirac spinor.
Moreover, it can only depend upon the
Lorentz-invariant quantities $p^2$, $(k^l)^2$ and $p\cdot k^l$, but the mass
shell conditions (\ref{v9b}) and (\ref{v12b2}) leave only
the 3-vectors $\bf p$ and $\bf k$ as independent variables.

The quantity $|\phi_l({\bf k},{\bf p})|^2$
is the probability density (homogeneous in space and time)
of destroying a nucleon (or creating an antinucleon)
with 4-momentum $k_l$ in the nuclear ground state
leaving the residual nucleus in the
quantum state $|{\bf p}-{\bf k},(A-1)_l>$.
Notice also that eqs. (\ref{v12a1}) to (\ref{v12b2}) hold valid
in the nonrelativistic limit, where the center-of-mass can be
correctly defined and its motion
accounted for. If however the center-of-mass is not properly singled out, as is
the case, e.g., of the nuclear shell model, then obviously the invariance for
space translations is lost.

Let us now illustrate how the energy of the nucleus and the momentum
distribution of its constituents can be expressed in terms of
$\phi_l({\bf k},{\bf p})$.

To start with, notice that
\begin{eqnarray}
\lefteqn{\sum_{l}\intt{k}\left|\phi_l({\bf k},{\bf
p})\right|^2=}\label{v13}\\
&&=\frac{1}{<{\bf p},A|{\bf p},A>}\int d^3y\,<{\bf p},A|\psi^\dagger({\bf
y})\psi({\bf y})|{\bf p},A>=A+4\Omega\intt{k}\nonumber
\end{eqnarray}
where the term proportional to the normalization volume $\Omega$ arises
from the Dirac sea and would be absent had we employed a
suitably defined normal product.

Furthermore, from (\ref{v12b2}) and using closure, one gets:
\begin{eqnarray}
\lefteqn{\sum_{l}\intt{k}\,\varepsilon_l({\bf k},{\bf p})
\left|\phi_l({\bf k},{\bf p})\right|^2}\label{v14}\\
&&=\frac{1}{<{\bf p},A|{\bf p},A>}\int d^3y\,
<{\bf p},A|\psi^\dagger({\bf y})\left(p_0-{H}\right)\psi({\bf y})
|{\bf p},A>\nonumber
\end{eqnarray}
which, exploiting the commutation relations for the field $\psi$,
can be recast in the form
\begin{eqnarray}
\lefteqn{\sum_{l}\intt{k}\,\varepsilon_l({\bf k},{\bf p})
\left|\phi_l({\bf k},{\bf p})\right|^2}\label{v14a}\\
&&=\frac{1}{<{\bf p},A|{\bf p},A>}<{\bf
p},A|\left({H}^0_N+{H}'\right)|
{\bf p},A>\;.\nonumber
\end{eqnarray}
The above relation
expresses the energy of the system associated with the fermionic
degrees of freedom. This quantity can of course be experimentally
inferred from the nuclear separation energy.
The "total" energy of the
nucleus, which is also experimentally known, reads instead
\begin{eqnarray}
\lefteqn{p_0=\sum_{l}\intt{k}\,\varepsilon_l({\bf k},{\bf p})
\left|\phi_l({\bf k},{\bf p})\right|^2}\label{v15}\\
&&+\int\frac{d^3k}{2\omega_k(2\pi)^3}
\frac{<{\bf p},A|\omega_k c^\dagger_kc_k|{\bf p},A>}{\nd}
-<0|{H}|0>\nonumber\quad.
\end{eqnarray}
One thus sees that in principle the possibility is offered, within
our hamiltonian model, to estimate the contribution to $p_0$ arising
from the mesonic components in the nuclear ground state. This is
expected to be small: should not this be the case then
the mesonic field expansion (\ref{v5})
would not be compatible with the normal product introduced in
(\ref{v2b'}).

Let us now somewhat generalize equation (\ref{v15}).
For this purpose we consider again the
baryon contribution to the total energy of the nucleus, introducing the
analogous of the spectral function in the nonrelativistic theory,
namely
\begin{equation}
S(k,{\bf p})=2\pi\sum_{l}\left|\phi_l({\bf k},{\bf
p})\right|^2\delta(k_0-\varepsilon_l({\bf k},{\bf p}))-S_{\rm vac}(k)
\label{v15n}
\end{equation}
which, as it is well known, is connected to the correlation function via a
Fourier transform
\begin{equation}
S(k,{\bf p})=
\frac{1}{2p_0}\Tr \int d^4y \,<{\bf p},A|
\psi^\dagger(y)\psi(0)|{\bf p},A>e^{-ik\cdot y}-S_{\rm vac}(k)
\label{v16n}
\end{equation}
and is crucial for expressing the $(e,e^\prime p)$
phenomenology\cite{PaSa-82}.

In term of the above, and reminding the need of subtracting the vacuum
contribution
\begin{equation}
S_{\rm vac}(k)=8\pi\Omega\delta(k_0+\epsilon_k)\label{v17d}\quad,
\end{equation}
the baryonic part of the energy can then be expressed as
follows:
\begin{equation}
p_0^{\rm nucl}=\intq{k} k_0 S(k,{\bf p})\label{v17n}\quad.
\end{equation}
The mesons contribution can be similarly handled by setting
\begin{equation}
S^0_\sigma(k,{\bf p})=2\pi \frac{<{\bf p},A|c^\dagger_kc_k|{\bf
p},A>}{2\omega_k
\nd}\delta(k_0-\omega_k)\label{v17c}
\end{equation}
in terms of which the mesonic energy reads
\begin{equation}
p_0^{\rm mes}=\intq{k}k_0 S^0_\sigma(k,{\bf p})\quad.
\end{equation}

The previous result (\ref{v15})
may then be generalized by considering the total
4-momentum, whose expression
\begin{equation}
p_\mu=\intq{k}k_\mu\left\{S(k,{\bf p})+S^0_\sigma(k,{\bf p})
\right\}\label{v17a}
\end{equation}
immediately follows from the above definitions. In (\ref{v17a})
$S$ and $S^0_\sigma$ can clearly be interpreted as the 4-momentum
distributions for nucleons and mesons respectively.

Note the asymmetric nature of  (\ref{v17a}):
$S(k,{\bf p})$, as defined in eq. (\ref{v15n}),
embodies the exact spectrum of the system, whereas
$S_\sigma^0(k,{\bf p})$ contains the free meson spectrum.
Thus $S(k,{\bf p})$ provides all the needed informations
about the interacting system: what is left out is the hamiltonian
of the free mesons. Of course one could have started by
defining an exact spectral function for the mesons, which would then contain
the full spectrum of the system: then a free spectral function for the
nucleons would be left out.

An energy integration leads to the 3-momentum distribution of the
fermions
\begin{equation}
n({\bf k},{\bf p})=\sum_{l}|\phil|^2
\label{v15b}
\end{equation}
and of the mesons
\begin{equation}
n_\sigma({\bf k},{\bf p})=\frac{<{\bf p},A|c^\dagger_kc_k|{\bf p},A>}
{2\omega_k<{\bf p},A|{\bf p},A>}\label{v16}
\end{equation}
respectively, so that the total momentum $\bf p$ of the nucleus will be
\begin{equation}
{\bf p}=\intt{k}{\bf k}\left\{n({\bf k},{\bf p})+n_\sigma({\bf k},{\bf p})
\right\}\quad.\label{v15c}
\end{equation}

Finally, note that $S$, $S^0_\sigma$ and $S_{\rm vac}$ are
Lorentz-covariant and the normalization of $S$ is
\begin{equation}
\intq{k}S(k,{\bf p})=\intt{k}n({\bf k},{\bf p})=A\quad.
\end{equation}
Clearly
\begin{equation}
\intq{k}S^0_\sigma=\intt{k}n_\sigma({\bf k},{\bf p})=N_\sigma
\end{equation}
provides the average number of mesons inside the nucleus.

\section{A Simple Case\label{sec3}}

In the previous section we have outlined a Lorentz-invariant
scheme for treating the atomic nucleus. In practical cases however
one is often forced to resort to a single-particle approximation
(Dirac-Hartree-Fock
approximation\cite{HoSe-81,CeSh-81}). In so doing
translational invariance is generally lost and the mesonic degrees of
freedom are dropped out of the problem.

Here we examine, in a
simple case, how a description of the nucleus solely in terms of nucleonic
degrees of freedom can be worked out respecting
Lorentz and translational invariance.

For this purpose we start by writing the purely nucleonic state vector
\cite{Di-49}
(sometimes also called Fock's line)
\begin{eqnarray}
\label{v19}
\lefteqn{|{\bf p},A>_F=(2\pi)^3\sqrt{\frac{2p_0}{A!}}}\\
  &&\int\prod_{i=1}^{A}\dki
\frac{m}{\epsilon_{k_i}}\psi_{\bfm{\scriptstyle p}}(k_1,\dots,k_A)
a^\dagger_{k_1}\dots a^\dagger_{k_A}|0>\delta({\bf p}-\sum_{i=1}^{A}
{\bf k}_i)\nonumber
\end{eqnarray}
where ${\bf k_i}\equiv({\bf k}_i,s_i)$, $s_i$ embodies the spin-isospin
variables, $a^\dagger_{k_i}$ and $a_{k_i}$ are the nucleon creation and
annihilation operators, $\epsilon_k=\sqrt{{\bf k}^2+m^2}$ and the vacuum
definition
\begin{equation}
a_{ks}|0>=0\qquad\qquad b_{ks}|0>=0\label{v7}
\end{equation}
is assumed to hold.

With the normalization (\ref{v9c}) for the states $|{\bf p},A>$,
and owing to (\ref{v6n}), the normalization of (\ref{v19})
reads
\begin{equation}
(2\pi)^3\int\prod_{i=1}^{A}\dki
\frac{m}{\epsilon_{k_i}}\left|\psi_{\bfm{\scriptstyle
p}}(k_1,\dots,k_A)\right|^2
\delta({\bf p}-\sum_{i=1}^{A}{\bf k}_i)=1
\label{v20}
\end{equation}
where $\psi_{\bfm{\scriptstyle p}}({\bf k})$ is also fully antisymmetric
in its arguments.

The advantage of (\ref{v19}) is of rendering straightforward the
connection with the non relativistic theory: indeed to recover the
latter it is sufficient to replace
$\psi_{\bfm{\scriptstyle p}}(k_1,\dots,k_A)$ with, e.g., a shell model
wave function in momentum space.

In addition (\ref{v19}) allows us to express the energy of the nucleus
and the 4-momentum
distribution of the nucleons in terms of the 3-momentum distribution only.
In fact since the 3-momentum distribution reads
\begin{equation}
n_s^F({\bf k},{\bf p})={m\over \epsilon_k}
\frac{_F<{\bf p},A|a^\dagger_{ks}a_{ks}|{\bf p},A>_F}{_F\nd_F}
\label{v1x}
\end{equation}
($s$ being the spin index), the energy will be given by
\begin{eqnarray}
p_0^F&=&\frac{_F<{\bf
p},A|\intt{k}\frac{m}{\epsilon_k}\epsilon_ka^\dagger_{ks}a_{ks}
|{\bf p},A>_F}{_F<{\bf p},A|{\bf p},A>_F}\nonumber\\
&=&\sum_{s}\intt{k}\epsilon_k
n_s^F({\bf k},{\bf p})\label{v21}\;.
\end{eqnarray}
Note that the Lorentz-invariant 3-momentum distribution (\ref{v1x}) is
normalized according to
\begin{equation}
\sum_{s}\intt{k}\frac{m}{\epsilon_k}
n_s^F({\bf k},{\bf p})=A\label{v23}\;,
\end{equation}
as it easily follows from (\ref{v20}), and that the 4-momentum distribution
is expressed in terms of the latter as follows
\begin{equation}
S^F(k,{\bf p})=2\pi\sum_{s} n_s^F({\bf k},{\bf p})
\delta(k_0-\epsilon_k)+8\pi\Omega\delta(k_0+\epsilon_k)\label{v24}
\end{equation}
(of course $S_\sigma^0=0$).

Also, for sake of completeness, we write down
the expressions for $n^F_s$ and $p_0^F$ in terms of the function
$\psi_p(k)$:
\begin{equation}
n_s^F({\bf k},{\bf p})=
(2\pi)^3A\int\prod_{i=2}^{A}\frac{d^3k_i}{(2\pi)^3}
{m\over \epsilon_k}\left|\psi_p(k_s,k_2,\dots
,k_A)\right|^2\delta({\bf p}-{\bf k}-\sum_{j=2}^{A}
{\bf k}_j)\label{v22}
\end{equation}
and
\begin{equation}
p_0^F=
(2\pi)^3\int\prod_{i=1}^{A}\frac{d^3k_i}{(2\pi)^3}
{m\over \epsilon_k}\left|\psi_p(k_1,\dots
,k_A)\right|^2\epsilon_{k_1}\delta({\bf p}-\sum_{j=1}^{A}
{\bf k}_j)\label{v7n}\;.
\end{equation}

The problem with (\ref{v19}) is that it can never correspond to a bound
state. Indeed,
in the rest frame, the energy of a bound system clearly satisfies
the condition
\begin{equation}
\epsilon<mA\quad.\label{v25}
\end{equation}
On the other hand, from (\ref{v21}) and (\ref{v23}) it follows
\begin{equation}
p_0^F=\sum_{s}\intt{k}\epsilon_k n_s^F({\bf k},{\bf p})>
m\sum_{s}\intt{k}n_s^F({\bf k},{\bf p})=mA\;.
\end{equation}

The above result is obvious because the mesons do not intervene
when the expectation value of the hamiltonian is taken in the state
(\ref{v19}). However it points to the need of
inserting the mesons in the model, preserving,
possibly,  a structure like (\ref{v19})
for the nuclear state. To illustrate how this scope can be achieved in a simple
scheme, let us consider
an infinite Fermi gas at zero temperature in the rest frame.

Notice that the well known momentum distribution for
such a system, namely
\begin{equation}
n_s^F({\bf k})=\Omega\theta(\k-k)\label{v26}\quad,
\end{equation}
is not Lorentz-covariant.
Yet we are willing to sacrifice temporarily Lorentz covariance in order to
take
advantage of the transparency of the Fermi gas model.

We start then from the equations of motion
\cite{LeWi-74,LeWi-79}:
\begin{eqnarray}
\left(\square-m_\sigma^2\right)\sigma_{\rm cl}&=&g<\overline{\psi}\psi>\\
\left(i\gamma\cdot\partial-m-g\sigma_{\rm cl}\right)\psi&=&0\quad.
\end{eqnarray}
These, for a uniform system (the Fermi gas), simplify to
\begin{eqnarray}
\sigma_{\rm cl}&=&-\frac{g}{m_\sigma^2}<\overline{\psi}\psi>\label{v8n}\;,\\
\left(i\gamma\cdot\partial-\tilde m\right)\psi&=&0\label{v9n}\;,\\
\tilde m&=&m-\frac{g^2}{m_\sigma^2}<\overline{\psi}\psi>\label{v10n}\;.
\end{eqnarray}
Thus an effective mass, always lower than the
bare one no matter which sign the coupling constant has, naturally appears.

We then perform the  canonical transformation
\begin{equation}
\sigma\to\sigma-\sigma_{\rm cl}\label{v11n}
\end{equation}
to get rid of the average value $\sigma_{\rm cl}$: this amounts to
replace the free nucleons and antinucleons with dressed ones. These
may be viewed as baryons carrying a cloud of $\sigma$-mesons, but
otherwise unaffected by the interaction according to eq. (\ref{v9n}).
To evaluate the energy of the system we accordingly use the field
representation
\begin{equation}
\psi({\bf y})=\sum_{s}\intt{k}\frac{\tilde m}{\tilde\epsilon_k}\left(
\tilde u_{ks}\alpha_{ks}e^{i\bfm{\scriptstyle k}\cdot\bfm{\scriptstyle y}}
+\tilde v_{ks}\beta^\dagger_{ks}
e^{-i\bfm{\scriptstyle k}\cdot\bfm{\scriptstyle y}}\right)\label{v12n}
\end{equation}
where the operators $\alpha$ and $\beta$ create and annihilate dressed
nucleons and antinucleons ($\tilde\epsilon_k=\sqrt{{\bf k}^2+\tilde m^2}$).
Proceeding now  along the previously outlined scheme,
but utilizing the representation (\ref{v12n}), which entails a
vacuum different from (\ref{v7}) and, consequently, a different
subtraction, we get
\begin{eqnarray}
p_0&=&\frac{1}{<{\bf p},A|{\bf p},A>}\sum_{s}\intt{k}
\tilde\epsilon_k\frac{\tilde m}{\tilde\epsilon_k}
<{\bf p},A|\alpha^\dagger_{ks}\alpha_{ks}+\beta_{ks}\beta^\dagger_{ks}
|{\bf p},A>\nonumber\\
&+&\mezzo\Omega m^2_\sigma\sigma^2_{\rm cl}
-4\Omega\intt{k}\tilde\epsilon_k\;.\label{v13n}
\end{eqnarray}
Here the first term has the lower bound $\tilde mA<mA$, whereas the second
one arises from the classical meson field via the canonical
transformation (\ref{v11n}). Thus a bound nuclear state becomes possible,
depending
upon the value of the coupling constant $g$ and of the nuclear density.
The last term in (\ref{v13n}) is again due to the lack of a suitably defined
normal product and its structure reflects the
presence of two vacua, one associated with the free particles and the
other with the dressed ones.

\section{General case\label{sec4}}

In this section we shall examine how
a many-body theory for a finite system can be cast in a relativistic
framework still preserving Poincar\'e invariance.

\subsection{The Single-particle Green's Function}

We start from the single-particle Green's function
\begin{equation}
G_{\bfm{\scriptstyle p}}(y,y')=-i
\frac{<{\bf p},A|T\left\{\psi(y),\psi^\dagger(y')\right\}|{\bf
p},A>}{\nd}
\label{v28}
\end{equation}
which obeys the Dyson's equation
\begin{equation}
\left(i\gamma\cdot\partial-m\right)G_{\bfm{\scriptstyle p}}(y,y')+\int d^4z
\Sigma_{\bfm{\scriptstyle p}}(y,z)G_{\bfm{\scriptstyle p}}(z,y')
=\gamma_0\delta^4(y-y')\quad.\label{v14n}
\end{equation}

Since we employ momentum eigenstates $G_{\bfm{\scriptstyle p}}$
can only depend upon $y'-y$ (translational invariance).
It will carry however an index $\bf p$ representing the
total momentum of the nucleus (the same applies to the self-energy
$\Sigma_{\bfm{\scriptstyle p}}$ as well).
Accordingly, we can work in the 4-momentum
space, where eq. (\ref{v14n}) becomes
\begin{equation}
\left(\gamma\cdot k-m-\Sigma_{\bfm{\scriptstyle
p}}(k)\right)G_{\bfm{\scriptstyle p}}(k)=\gamma_0
\label{v29}\quad.
\end{equation}

Obviously, eqs. (\ref{v14n}) and (\ref{v29}) are meaningful only if a
prescription is given for calculating
the self-energy. This is far from being trivial in a relativistic context.
A perturbative approach will not
work, since a bound state of a finite assembly of nucleons cannot be obtained
from a non interacting system via the summation of Feynman diagrams.
Rather a collection of nucleons bound in an external (possibly
self-consistent) well (as in refs.
\cite{HoSe-81,CeSh-81}) might be taken as a starting point.
This however entails to give up Lorentz and translational invariance
from the very beginning in the hope that the summation of appropriate
classes of perturbative diagrams will restore those
symmetries. If this turns out to be the case,
then eqs. (\ref{v14n}) and (\ref{v29}) are meaningful.

Here our
standpoint is that $\Sigma_{\bfm{\scriptstyle p}}$ exists, is calculable,
Lorentz-covariant and transforms like a $\gamma$-matrix.
As a consequence $G_{\bfm{\scriptstyle p}}$ will exist as well. However
to write down a Dyson's equation of the type
\begin{equation}
G_{\bfm{\scriptstyle p}}=G_{\bfm{\scriptstyle p}}^0
+G_{\bfm{\scriptstyle p}}^0\Sigma_{\bfm{\scriptstyle p}}
 G_{\bfm{\scriptstyle p}}\label{v3x}\;,
\end{equation}
preserving at the same time the  Poincar\'e invariance together with
the existence of a bound nuclear state, turns out to be a task beyond reach.

Therefore, we start from the spectral representation of
$G_{\bfm{\scriptstyle p}}$ itself, namely
\begin{equation}
G_{\bfm{\scriptstyle p}}(k)=\frac{2p_0}{\nd}\left\{\sum_{l}
\frac{\phil\phil^*}{k_0-\ve
-i\eta}+\sum_{l}\frac{\chil\chil^*}{k_0-\tilde\ve
+i\eta}\right\}\label{v30}\;,
\end{equation}
which introduces,
in parallel to the previously discussed $\phil$, a
new function
\begin{equation}
\chil=\tilde\cl<{\bf p}+{\bf k},(A+1)_l|\psi^\dagger(0)|{\bf p},A>\;,
\end{equation}
with
\begin{equation}
\tilde\cl=\sqrt{\frac{M_l^{A+1}}{2p_0\epsilon_l^{A+1}({\bf p}+{\bf k})}}\;,
\label{v4x}
\end{equation}
whose modulus square yields the probability of creating  a nucleon (or
destroying an antinucleon)
with 4-momentum $k_l$ on the nuclear ground state $|{\bf p},A>$
leaving the residual system in the quantum state $|{\bf p}+{\bf k},(A+1)_l>$.
Moreover, $\epsilon^{A+1}_l({\bf p+k})$ and
$\tilde\ve$
refer to an excited state with baryonic number $A+1$ and are defined in analogy
with the $A-1$ case.

It helps, in grasping the significance of the singularities in the
propagator, to think to the infinite Fermi gas where one has
\begin{eqnarray}
\lefteqn{G_{\bfms{p}=0}^F(k)={\slash{k}+m\over 2\sqrt{k^2+m^2}}\left\{
\theta(k-\k)\over k_0-\sqrt{k^2+m^2}+i\eta\right.}\\
&&\left.+{\theta(\k-k)\over k_0-\sqrt{k^2+m^2}-i\eta}
-{1\over k_0+\sqrt{k^2+m^2}-i\eta}\right\}\nonumber\;.
\end{eqnarray}
Here the antinucleon states are entirely embodied in the last term of
the rhs. This separation however becomes impossible when the
interaction is introduced.
Accordingly in (\ref{v30}) the antinucleons are actually embodied in
both terms of the rhs (not only in the first).

To better understand the above spectral representation
it might help to think to a system in which either a
nucleon has been annihilated or an antinucleon has been created. This
picture strictly speaking
is not correct, since the nucleus is actually described by
a superposition of states
with different number of nucleons and antinucleons; we can however
start by considering an assembly of noninteracting nucleons in the rest frame
and then annihilate a
nucleon or create an antinucleon. In this way two sets of states are
obtained, denoted, with self-explaining notation, $(A-1N)$ and $(A+1\tilde N)$
respectively. Then switching  on the interaction we generate two classes of
physical states, still somewhat linked to the original $(A-1N)$ and
$(A+1\tilde N)$ ones.
The positive energy solutions are
associated to the former, the negative energy ones to the latter, both
stemming from the singularities of the first term on the rhs of
(\ref{v30}). The singularities of the second term relate to $(A+1N)$
states, i.e. states in which a $N\overline{N}$ pair exists.

Of course
\begin{eqnarray}
-i\Tr\intq{k}\G&=&\frac{2p_0}{\nd}\sum_{l}\intt{k}
\left|\phil\right|^2\nonumber\\
&=&\frac{A}{\Omega}+4\intt{k}\label{v31}\;.
\end{eqnarray}

By inserting then (\ref{v30}) in (\ref{v29}) one obtains the following
eigenvalue equation for $\phil$ in terms of the $\bfm \alpha,\beta$
representation for the Dirac's matrices:
\begin{eqnarray}
\lefteqn{G_\bfms{p}^{-1}(k)\phil=}\label{v32}\\
&&=\left(\ve-\bfm\alpha\cdot {\bf k}-\beta m-\beta\Sigma_p(k)\Bigm|_{k_0
=\ve}\right)\phil=0\;.\nonumber
\end{eqnarray}
An analogous equation holds as well for the $\chil$.

In (\ref{v32}) the eigenvalue $\ve$, as told by eq. (\ref{v12b2}),
represents the difference between the ground state energy of the
system in the state $|{\bf p},A>$ (moving with total momentum $\bf p$) and
the energy of the system in the  state $|{\bf p}-{\bf k},(A-1)_l>$
(moving with momentum $\bf p-k$ and intrinsically left in the quantum state
$l$). It can be regarded as the energy of the quasi-hole. On the same
footing $\tilde\ve$ represents the energy of a quasi-particle.

\subsection{The Self-Energy and the Final State Interaction}

The crucial quantity in the many-body equation for the
spectral amplitude is clearly the self-energy.
In this subsection we shall obtain the latter from the
interaction between a nucleon and the daughter nucleus, thus linking the
self-energy to the so-called final-state interaction. An attempt in the
same direction has been performed in ref. \cite{CeSh-86-B}.

Let us introduce a new field operator $\ppsi{Y}$ associated with the
$A-1$ nucleus viewed as a composite fermion which can exist in
different internal states:
\begin{equation}
\ppsi{Y}=\sum_{l}\intt{p}\frac{M_l^{A-1}}{\epsilon_l^{A-1}({\bf p})}
U_{l\bfm{\scriptstyle p}}({\bf Y}){\cal A}_{l\bfm{\scriptstyle p}}\;.
\end{equation}
Here the operator ${\cal A}_{l\bfm{\scriptstyle p}}$ annihilates a
nucleus of baryonic number $A-1$ in the excited state $l$ and with total
3-momentum $\bf p$ (we neglect, as irrelevant in the energy range
of actual interest, the
component of the field that creates the corresponding antinucleus).
$U_{l\bfm{\scriptstyle p}}({\bf Y})$
is a Dirac spinor, normalized in Fourier transforms according to
\begin{eqnarray}
U_{l\bfm{\scriptstyle p}}({\bf Y})&=&U_{l\bfms{p}}e^{i\bfms{Y}\cdot
\bfms{p}}\nonumber\\
U^\dagger_{l\bfms{p}}U_{l'\bfms{p}}&=&\frac{\epsilon_l^{A-1}({\bf p})}{
M_l^{A-1}}\delta_{ll'}\quad.\label{eq76j}
\end{eqnarray}
Next we introduce the four-point Green's function
\begin{equation}
\GG(y,Y;y',Y')=\frac{<{\bf p},A|T\left\{\psi(y)\Psi(Y)
\psi^\dagger(y')\Psi^\dagger(Y')\right\}|{\bf p},A>}{\nd}
\end{equation}
which obeys the following Bethe-Salpeter-like equation
\begin{eqnarray}
&&\GG(y,Y;y',Y')=\GG^0(y,Y;y',Y')-i\int
d^4y_1\,d^4y_2\,d^4Y_1\,d^4Y_2\nonumber\\
&&\times \GG^0(y,Y;y_1,Y_1){\cal V}_{\bfms{p}}
(y_1,Y_1;y_2,Y_2)\GG(y_2,Y_2;y',Y')
\label{eq81v}
\end{eqnarray}
where ${\cal G}^0_{\bfms{p}}$ is the propagator
of the $A^{\rm th}$ nucleon and of the $(A-1)$ daughter nucleus in the absence
of their mutual interaction
${\cal V}_{\bfms{p}}$.

Since all the functions $\GG$, $\GG^0$ and ${\cal V}_{\bfms{p}}$
carry a definite total momentum $\bf p$, they
are best handled in Fourier transforms: accordingly let us set
\begin{eqnarray}
\lefteqn{{\cal F}(k,k';p)=\int d^4y\,d^4y'\,d^4Y\,d^4Y'}\\
&&e^{ik\cdot y}e^{-ik'\cdot y'}e^{i(p-k)\cdot Y}e^{-i(p-k')\cdot
Y'}{\cal F}_{\bfms{p}}(y,Y;y',Y')\nonumber
\end{eqnarray}
where ${\cal F}$ can be ${\cal G}$, ${\cal G}^0$ or ${\cal V}$.
In this representation
${\cal G}^0$ reads
\begin{equation}
{\cal G}^0(k,k';p)=(2\pi)^4\delta(k-k')g^0(k)G^0_{A-1}(p-k)\;.
\end{equation}
Here $g^0(k)$ is the relativistic free nucleon propagator in the vacuum
(not to be confused, of course, with the $G^0$ appearing in eq. (\ref{v3x})).
Indeed
the latter is ill-defined, whereas the former (disregarding
renormalization effects) reads
\begin{equation}
\left[g^0(k)\right]^{-1}=k_0-\bfm\alpha\cdot{\bf k}-\beta m\;\;.
\end{equation}
The relativistic free propagator for the $(A-1)$ nucleus
(not interacting with the $A^{th}$ nucleon), is given by
\begin{equation}
G_{A-1}^0(p)=\sum_{l}\frac{M_l^{A-1}}{\epsilon_l^{A-1}({\bf p})}
\frac{U_{l\bfms{p}} U_{l\bfms{p}}^\dagger}{p_0-\epsilon_l^{A-1}({\bf
p})+i\eta}\;.
\end{equation}
The Fourier transformed Bethe-Salpeter equation reads then
\begin{eqnarray}
\lefteqn{{\cal G}(k,k';p)=(2\pi)^4g^0(k)G^0_{A-1}(p-k)\delta(k-k')}
\label{eq78bis}\\
&&+g^0(k)G^0_{A-1}(p-k)\int d^4k{\cal V}(k,k;p)
{\cal G}(k,k';p)\nonumber
\end{eqnarray}
We introduce now the Bethe-Salpeter wave function
in line with the usual convention \cite{GeLo-51,ZuTj-80,ZuTj-81}

\begin{equation}
\varphi_{\bfms{p}}(y,Y)=-i<0|T\left\{\psi(y)\Psi(Y)\right\}|
{\bf p},A>=\intq{k}
\varphi_{\bfms{p}}(k)e^{-iky-i(p-k)Y}\label{eqn80}
\end{equation}
whose spectral representation reads
\begin{eqnarray}
\lefteqn{\varphi_{\bfms{p}}(k)=}\label{eq83v}\\
&&(2\pi)^3<0|\psi(0)\frac{\delta({\bf k}-
\hat{{\bf P}})}{k_0-H+i\eta}\Psi(0)+\Psi(0)
\frac{\delta({\bf k}-{\bf p}-\hat{{\bf P}})}{
k_0-p_0+H-i\eta}\psi(0)|{\bf p},A>\nonumber\;,
\end{eqnarray}
$\hat{\bf P}$ being the 3-momentum operator and
$p_0=\sqrt{M^2_A+{\bf p}^2}$ both in eqs. (\ref{eqn80}) and (\ref{eq83v}).

{}From (\ref{eq78bis}) an equation for $\varphi_{\bfms{p}}(k)$
is then deduced by writing  down the Lehmann representation of
${\cal G}(k,k';p)$ and selecting  those contributions having the vacuum as
intermediate state. This scope is achieved by integrating over $p_0$ around a
small circle surrounding the point $p_0=\sqrt{{\bf p}^2+M_A^2}$ which indeed
corresponds to the pole of the propagator having the vacuum as
an intermediate state. Thus we find
\begin{equation}
\oint\frac{dp_0}{2\pi i}{\cal G}(k,k';p)=\varphi^*_{\bfms{p}}(k)
\varphi_{\bfms{p}}(k')
\end{equation}
and since ${\cal G}^0$ is regular at $p_0=\sqrt{{\bf p}^2+M^2_A}$ one
gets, by integrating eq. (\ref{eq78bis}) over $p_0$,
\begin{equation}
\varphi_{\bfms{p}}(k)=- i\oint\frac{dp_0}{2\pi i}
g^0(k)G^0_{A-1}(p-k)
\intq{k'}{\cal V}(k,k';p)\varphi_{\bfms{p}}(k')\label{eqnstar}\;.
\end{equation}

Remarkably, the Bethe-Salpeter wave function $\varphi_{\bfms{p}}(k)$ turns
out to be connected with the spectral amplitude $\phil$. In fact, as seen from
eq. (\ref{eq83v}), $\varphi_{\bfms{p}}(k)$
is  a meromorphic function in the variable $k_0$, with poles located at
the energies of the free baryons (first term of the rhs of eq.
(\ref{eq83v})~) and at the energies of the quasiparticle (second term).
We can thus insert a complete set of eigenstates of the $(A-1)$ system
in the second term of the rhs of eq. (\ref{eq83v}), integrate over a
small circle around the selected pole and, by reminding the definition
(\ref{v12a2}) of $\phi_l$, we find
\begin{equation}
\oint\frac{dk_0}{2\pi i}\varphi_{\bfms{p}}(k)=
\left({M^{A-1}_L\over \epsilon_l^{A-1}({\bf p-k})}\right)U_{l
\bfm{\scriptstyle p}-\bfm{\scriptstyle k}}\frac{1}{\cl}\phil\;.
\end{equation}
{}From the above, using (\ref{eq76j}) and (\ref{eq83v}), the connection we are
searching for, namely
\begin{equation}
\phil=\cl\oint\frac{dk_0}{2\pi i}U^\dagger_{l(
\bfms{p}-\bfms{k})}\varphi_{\bfms{p}}
(k)
\end{equation}
follows.

Next we act on (\ref{eqnstar}) with $[g^0]^{-1}$ and
again select the poles by integrating over $k_0$ around the $l^{\rm th}$
eigenvalue of the $(A-1)$ system getting
\begin{eqnarray}
&&\left(\ve-\bfm{\alpha}\cdot{\bf k}-\beta m\right)\phil=\\
&&\qquad\qquad\cl
\intq{k'}U^\dagger_{l(\bfms{p}-\bfms{k})}{\cal V}(k_l,k';p)
\varphi_{\bfms{p}}(k')\nonumber\;.
\end{eqnarray}

The above equation provides a tool for calculating microscopically the
self-energy: indeed given an effective nucleon-nucleus interaction ${\cal
V}(k,k';p)$ the Bethe-Salpeter wave function $\varphi_{\bfms{p}}(k)$
can be found and the product of the baryon self-energy with the spectral
amplitude $\phil$ can be determined.

Note also that the finite size of the system is actually embedded in ${\cal
V}(k,k';p)$.

Clearly for  ${\cal V}(k,k';p)$ a model is needed. Interestingly,
if we choose the expression
\begin{equation}
{\cal V}(k,k';p)=(2\pi)^3\delta({\bf k}-{\bf k}')(\beta\sigma+\omega)
\label{vz2.1}
\end{equation}
for the effective interaction, then
the well known self-energy of the QHD, namely
\begin{equation}
\Sigma_\bfms{p}(k)=\sigma+\beta \omega\;,
\end{equation}
is recovered, $\sigma$ and $\omega$ being the standard scalar and vector
fields of QHD.

\subsection{The Structure of the Self-Energy}

Having examined a microscopical model for the self-energy and its
link with the final state interaction, we now write down its general
expression, namely
\begin{equation}
\Sigma(k^2,k\cdot p)=S+\gamma_\mu
V^\mu+{1\over 2M_A}\left[\gamma_\mu,\gamma_\nu\right] T^{\mu\nu}\;,
\label{vself}
\end{equation}
which fulfills the constraints imposed by Poincar\'e covariance.
Our purpose is to explore the predictions that can be made on the relativistic
quasiparticle spectrum without specifying the details of the baryon
self-energy.
In (\ref{vself})
\begin{equation}
V^\mu=k^\mu V+p^\mu V'\label{v20n}
\end{equation}
and
\begin{equation}
T^{\mu\nu}=k^\mu p^\nu T
\end{equation}
are a Lorentz vector and a second rank tensor respectively, whereas
$S$, $V$ and $V'$ are scalars, in general complex,
whose dependence is  upon the invariants
$k^2$ and $k\cdot p$. Alternatively, we can choose as independent variables,
$k^2$ and $(p-k)^2$: this might be more convenient as
it will be discussed in sect. \ref{subs4.5}.

Our expression for $\Sigma_\bfms{p}$ is similar to the one of ref.
\cite{Mi-75}, often considered in the literature, but we remind that
translational invariance
strongly constrains its structure. We also remind that terms
with $\gamma_5$ are neglected as parity-violating;
concerning the impact of other discrete
symmetries on $\Sigma_\bfms{p}$ we refer the reader to \cite{Mi-75}.

With the self-energy (\ref{vself}) the inverse propagator entering eq.
(\ref{v32}) reads
\begin{equation}
G_{\bfms{p}}^{-1}(k)=(1-V)\left\{k_0-vp_0-\bfm\alpha\cdot({\bf k}-v{\bf p})
-\beta m^*-{t\over 2M_A}\beta\left[\slash{k},\slash{p}\right]\right\}
\label{vz1.1}
\end{equation}
where the definitions
\begin{eqnarray}
m^*&=&{m+S\over 1-V}\;,\\
v&=&{ V'\over 1-V}\;,
\end{eqnarray}
and
\begin{equation}
t={ T\over 1-V}
\end{equation}
have been introduced.
The above Green's function $G_{\bfms{p}}$ describes a moving quasiparticle
interacting with the daughter nucleus.
However, before discussing the physical significance of the functions
introduced so far, namely $m^*$, $v$, $t$ and $V$,
we first need to find the poles of the Green's function, i.e. to solve
the eigenvalue equation for the spectral amplitude $\phi_l$, and then to
study the analytical properties of $\Sigma_\bfms{p}$.

\subsection{Solution of the Eigenvalue equation}

To solve eq. (\ref{v32}) we look for the poles
of the propagator
\begin{eqnarray}
\lefteqn{G_\bfms{p}(k)={1\over (1-V)\left(k^2-m_1^2+i\eta\right)}}
\label{vz1.2}\\
&&\left\{k_0-vp_0-\bfm\alpha\cdot({\bf k}-v{\bf p})+\beta m^*-{t\over 2M_A}
\left[\slash{k},\slash{p}\right]\beta\right\}
\nonumber
\end{eqnarray}
where a new mass
\begin{equation}
m_1^2={m^*}^2+2vk\cdot p-v^2p^2+\left(k^2-\left({k\cdot p\over
M_A}\right)^2\right)t^2
\label{vz1.3}
\end{equation}
has been introduced. The small positive imaginary part in the
denominator is meant to be effective as far as $m^*$, $v$ and $t$ are real.

Clearly the poles of $G_\bfms{p}$ coincides with the eigenvalues $\ve$
of (\ref{v32}) and are the roots of the equation
\begin{equation}
k^2=m_1^2(k^2,k\cdot p)\quad.
\label{vz1.4}
\end{equation}
We shall discuss the regularity of the functions $m^*$, $v$ and
$t$ later on. In the regions where they are well behaved and $m_1^2$ is
positive definite
the nonlinear equation (\ref{vz1.4}) might display positive
\begin{equation}
\varepsilon_l^+=\sqrt{{\bf k}+m^2_1}\bigm|_{k_0=\varepsilon_l^+}
\label{vz1.5}
\end{equation}
and negative
\begin{equation}
\varepsilon_l^-=-\sqrt{{\bf k}+m^2_1}\bigm|_{k_0=\varepsilon_l^-}
\label{vz1.6}
\end{equation}
solutions. Notice that these in the rest frame become
\begin{equation}
\varepsilon_l^\pm =M_Av_l\pm\sqrt{(1-t^2_l){\bf k}^2+{m^*_l}^2}
\label{vz1.7}
\end{equation}
where the suffix $l$ reminds that the various quantities have to be
evaluated at $k_0=\varepsilon^\pm_l$.

We thus see that $v_l$ yields an overall
frequency shift, whereas $m_l^*$ sets the splitting between the two
branches of the spectrum. Noteworthy is that an energy shift,
a mass renormalization and the introduction of an effective
momentum
\begin{equation}
{{\bf k}}_{\rm eff}=\sqrt{1-t^2}\,{{\bf k}}
\end{equation}
allow a recovering of the structure of the Dirac's equation
eigenvalues. In particular the scalar component of $\Sigma_\bfms{p}$
renormalizes the mass, the vector one partly renormalizes the Green's
function residua and partly shifts the spectrum and finally the tensor
term modifies the momentum.

Of course $m_1^2$ is not an even function of $k_0$
(due to its dependence upon $p\cdot k$), so that in general
$\varepsilon_l^+\not= -\varepsilon_l^-$. Also, the existence of one
solution does not imply, in general, the existence of the other one.

{}From the analogy with the Dirac's equation we can interpret these
solutions as describing the propagation of a nucleon (or of a hole,
depending upon the boundary conditions) or of an antinucleon in the
medium.

Finally we note that the positive branch is the meaningful one in the non
relativistic limit, where
\begin{equation}
\varepsilon_l({\bf k})\approx m^*_l+M_A v_l+{(1-t_l^2){\bf
k}^2\over 2m^*}\;.
\end{equation}

Now we search for the spinor structure of the spectral amplitude $\phi_l$.
The equation (\ref{v32}) can be recast, with the help of
(\ref{vz1.2}), in the  form
\begin{equation}
\left\{\ve-v_lp_0+\bfm\alpha({\bf k}-v_l{\bf p})+\beta m^*_l-{t_l\over 2M_A}
\left[\slash{k}\bigm|_{k_0=\ve},\slash{p}\right]\beta\right\}\phil=0
\label{vz1.8}
\end{equation}
which defines $\phil$ up to a normalization.

Now for positive frequency solutions we get
\begin{equation}
\phi_l^+({\bf k},{\bf p})=\sum_{s} \left(1+{\gamma\cdot p\over
M_A}t\right)u_s {\cal N}_{ls}^+({\bf k},{\bf
p})\Bigm|_{k_0=\varepsilon_l^+}
\label{vz1.9.0}
\end{equation}
where
\begin{eqnarray}
&&u_s=\sqrt{{\cal E}+{\cal M}\over 2{\cal M}}\left({\varphi_s\atop
{\bfm{\sigma}\cdot\bfm{{\scriptstyle\cal K}}\over {\cal E}+{\cal
M}}\varphi_s}\right)\label{vz1.9}\;,\\
&&\bfm{\cal K}={\bf k}(1-t^2)-{\bf p}\left(v-{m^*\over M_A}t-{k\cdot p
\over M_A^2}t\right)\label{vz1.10},\\
&&{\cal M}=m^*-{t\over M_A}\left(k\cdot p-vp^2\right)\label{vz1.11}
\end{eqnarray}
and
\begin{equation}
{\cal E}=\sqrt{{\cal K}^2+{\cal M}^2}\label{v1.12}\;.
\end{equation}
For negative frequency solutions we have instead
\begin{equation}
\phi_l^-({\bf k},{\bf p})=\sum_{s} \left(1+{\gamma\cdot p\over
M_A}t\right)v_s {\cal N}_{ls}^-({\bf k},{\bf
p})\Bigm|_{k_0=\varepsilon_l^-}\label{vz1.13}
\end{equation}
with
\begin{equation}
v_s=\sqrt{{\cal E}+{\cal M}\over 2{\cal M}}\left({
-{\bfm{\sigma}\cdot\bfm{{\scriptstyle\cal K}}\over {\cal E}+{\cal
M}}\varphi_s\atop \varphi_s}\right)\label{vz1.14}\;.
\end{equation}
In (\ref{vz1.9}) and (\ref{vz1.14}) $\varphi_s$ is the Pauli spinor
corresponding to the spin polarization $s$.
We normalize the spinors $u$ and $v$ according to
$\overline{u}u=1$, $\overline{v}v=-1$,
while the normalization coefficient $\cal N$ remains to be fixed.
This is done by observing that
near the pole $\varepsilon_l^{(\pm)}({\bf k},{\bf p})$
the Green's function behaves like
\begin{equation}
G_{\bfms{p}}\simeq\frac{2p_0}{<{\bf p}A|{\bf p}A>}\frac{\phi_l^\pm
({\bf k},{\bf p}){\phi^\pm_l}^\dagger({\bf k},{\bf p})}
{{k}_0-\varepsilon({\bf k},{\bf p})-i\eta}\;,\label{eq114}
\end{equation}
but on the other hand
\begin{equation}
G_\bfms{p}=\left\{k_0-\bfm\alpha\cdot{\bf k}-\beta
m-\beta\Sigma\right\}^{-1}\quad.\label{eq115}
\end{equation}
By comparing (\ref{eq114}) and (\ref{eq115}) near
the pole the relation
\begin{equation}
\frac{2p_0}{<{\bf p}A|{\bf p}A>}\left|\phi_l^\pm
({\bf k},{\bf p})\right|^2=\frac{1-{vp_0\over k_0}}{(1-V)\left(
1-{m_1\over k_0}\frac{\partial m_1}{\partial
k_0}\right)}\Biggm|_{k_0=\varepsilon_l^\pm}
\label{vz1.15}\;,
\end{equation}
which fixes the normalization of $\phi_l^\pm$, follows.

\subsection{Properties of the Self-Energy\label{subs4.5}}

In this subsection we first shortly
study the analytic properties of the self-energy
starting from its microscopic definition. For this purpose we consider
the equation of motion for the field operator $\psi(y)$
\begin{equation}
\left(i\gamma\cdot\partial-m-{\cal U}\right)\psi(y)=0
\label{vz1.16}
\end{equation}
where the operator ${\cal U}$ is defined through the equal time commutator
\begin{equation}
\left({\cal U}\psi\right)(y)=\left[\psi(y),H^\prime(y_0)\right]\;.
\label{vz1.17}
\end{equation}
The operator $\cal U$ here introduced is model dependent. For our
hamiltonian (\ref{v1}) it reads
\begin{equation}
({\cal U}\psi)(y)=g\sigma(y)\psi(y)\;.
\label{vz17.1.1}
\end{equation}
Now, by operating with the operator on the
lhs of (\ref{vz1.16}) on the Green's function
definition and by exploiting the Dyson's equation (\ref{v14n}), one
obtains
\begin{eqnarray}
\lefteqn{\left(\Sigma G\right)(y,y^\prime)\equiv\int d^4z
\Sigma(y,z)G(z,y^\prime)}\label{vz18.1}\\
&&=2ip_0{<A{\bf p}|T\left\{\left({\cal U}\psi\right)(y)
\psi^\dagger(y^\prime)\right\}|A{\bf p}>\over \nd}\nonumber\;.
\end{eqnarray}
{}From (\ref{vz18.1}) the
Lehmann representation
\begin{eqnarray}
\lefteqn{\Sigma(k)G(k)={2M_A\over<A|A>}\times}\label{vz18.1.1}\\
&&\left\{\sum_{l}{<A|\psi^\dagger(0)|-{\bf k},(A-1)_l>
<-{\bf k},(A-1)_l|({\cal U}\psi)(0)|A>\over k_0-\varepsilon_l({\bf k})-i\eta}
\right.\nonumber\\
&&\left.+\sum_{l}{<A|({\cal U}\psi)(0)|-{\bf k},(A+1)_l>
<-{\bf k},(A-1)_l|\psi^\dagger(0)|A>\over k_0-\tilde\varepsilon({\bf k})
+i\eta}\right\}\nonumber
\end{eqnarray}
is deduced {\em in the rest frame.}

Let us now restrict ourselves to the discrete spectrum only.
Here, as a function of $k_0$, $G(k)$  (which, strictly speaking, is a
distribution)
displays a pole in correspondence of each discrete
eigenvalue. Owing to (\ref{vz18.1.1})
so does $\Sigma G$, hence $\Sigma$ must be regular.
Furthermore a zero of $G$ certainly exists
between two neighboring poles, where instead $\Sigma G$ is
regular. Hence {\em $\Sigma$ must diverge where $G$
vanishes.} On the other hand
in the continuous spectrum region $G$ is
finite, nonvanishing and complex on the real axis: as a
consequence the same occurs for $\Sigma$.

In conclusion $\Sigma$, and consequently $m^*$, $v$, $t$ and
$V$, are well-behaved functions in the neighborhoods of any point of the
spectrum of $G$.
However they are singular elsewhere on the real axis (actually between
the poles of $G$), in correspondence to the discrete part of the spectrum
of the nucleus, whose occurrence is allowed by its finite size.

Now let us shortly comment the dependence of the eigenvalues upon $\bf
k$ and $\bf p$, which is fixed by eq. (\ref{v12b2}) (at least for the
advanced part of the spectrum; the generalization however is obvious):
\begin{equation}
\varepsilon_l({\bf k},{\bf p})=\sqrt{{\bf p}^2+M_A^2}
-\sqrt{({\bf p}-{\bf k})^2+(M^{A-1}_l)^2}\;.
\label{vz4.2}
\end{equation}
This relation suggests the choice of
$k^2$ and $(p-k)^2$ as independent variables for expressing
$m^*$, $v$, $t$ and $V$, because it entails
\begin{equation}
(p-k)^2=\left(M^{A-1}_l\right)^2\;.
\label{vz4.2.1}
\end{equation}

Now the eigenvalues satisfy
eq. (\ref{vz1.4}), which, setting $f=k^2-m_1^2$, can be
recast as follows:
\begin{equation}
f((p-k)^2,k^2)=0
\end{equation}
which in turn can be inverted to yield
\begin{equation}
(p-k)^2=\varphi_l(k^2)\;.
\label{vz4.2.2}
\end{equation}
Combining then (\ref{vz4.2.1}) with (\ref{vz4.2.2}) one finally gets
\begin{equation}
\varphi_l(k^2)=\left(M^{A-1}_l\right)^2\quad \forall k^2
\label{vz4.2.3}
\end{equation}
which constrains the functional dependence of the self-energy.

As a simple example fulfilling eq. (\ref{vz4.2.2}) we consider the case
$v=1$ and $t=0$ (which implies the absence of an effective momentum)
and requiring in addition to have $m^*$ depending upon $(p-k)^2$ only.
Then the eigenvalue equation (\ref{vz1.4}) reduces to
\begin{equation}
(p-k)^2={m^*}^2((p-k)^2)\;.
\label{vz4.1.5}
\end{equation}
In this simple example the only unconstrained  quantity left out is $V$, which
can then be fixed by comparing with the experimental data.

In conclusion the purpose of this subsection has been to point out how
the structure of the self-energy is constrained, in a Lorentz covariant
traslationally invariant framework for a finite system,  by the analytic
properties it must display and by its dependence upon the momentum $p$
of the nucleus.

\section{Conclusions}

In this paper we propose a relativistic, hamiltonian theory of nuclei,
which also fulfills translational
invariance.

In order to set up such a framework we focused our attention on
the spectral amplitude: indeed
by one side the latter represents a most suitable tool for extending
a basic concept of the nuclear physics in the non-relativistic
domain, namely the one of single particle orbital, to the relativistic
regime, on the other side it turns out to be natural to
express the energy of the nucleus and its 3- and 4-momentum
distributions, in a scheme respecting covariance and in which the number
of nucleons is not conserved, precisely in terms of the spectral
amplitude.

We achieve our scope within a model, namely assuming the nucleus to be
a composite system of nucleons and mesons. Although the latter are
not elementary degrees of freedom, yet we believe
that they can be considered
as  ``effective'' building blocks useful in describing the nuclear
structure in a quite wide kinematical
region: therefore to them we have first applied our formalism.
On the other hand if and when more fundamental degrees of
freedom, namely the quarks and gluons, should be accounted for, then our
theory can equally well be applied to this new dynamical situation.

As an aside, in sect. 3 of the present paper we have also discussed the
limiting case, opposite to the above one, where the energies involved
are so small that the nucleus lends itself to be described solely in
terms of (suitably renormalized) nucleonic
degrees of freedom. This is the extreme situation corresponding to the
more traditional nuclear physics where the mesonic degrees of freedom
are eliminated in favour of static forces acting between the nucleons.
Of course the limits of such a scheme are met when the
Meson Exchange Currents (MEC) intervene in a decisive manner, as, e.g., in the
electrodisintegration of the deuteron at large momentum transfer
\cite{FaAr-79,Ar-82}.

In the present paper we have not addressed the question related to the global
gauge invariance of the theory, a symmetry required to consistently
account for both the
currents and the forces carried by the mesons. We have rather centered
our attention on the dynamics of the many-body system exploring in
particular the fermionic self-energy,
the basic quantity entering into the definition of the fermion propagator.
More specifically we started from its relativistic, translational
invariant definition and by
combining its Lehmann representation with the relativistic
Dirac's equation we succeeded in deriving an equation obeyed by
the spectral amplitude $\phi_l$ itself.
This one can be solved once the self-energy
is known and the associated eigenvalues directly provide the energy spectrum
of the nucleus. We have actually done that assuming a general Poincar\'e
invariant  self-energy: thus we  obtained
the spectral amplitude and the spectrum
of the nucleus. Notably the latter turned out to be a shifted
Dirac's spectrum with an effective mass and momentum.

In attempting a more microscopic approach, in particular to account for
the final state interaction, we considered a Bethe-Salpeter-type
equation whose solution happens to be closely connected with the
spectral amplitude
$\phi_l$. Worth noticing is that our Bethe-Salpeter approach, for
a particular choice of the
nucleon-nucleus interaction and for infinite nuclear matter,
reduces to QHD.

Finally we have explored the
analytical properties of the self-energy in the energy variable. We have
found that, in correspondence of the zeros of the fermion propagator,
the self-energy displays simple poles, which are clearly connected with
the bound states of the discrete part of the nuclear spectrum. The
locations of the poles is dependent upon the momentum $\bf p$ of the
nucleus, which is perfectly defined in our translational invariant
framework. The possibility is thus offered of exploring how the spectrum
of the nucleus modifies as a function of $\bf p$: this seems to us a
fascinating topic to investigate, in particular we conjecture that in
the infinite momentum frame the nuclear spectrum will considerably
simplify.

In conclusion in the present paper we have outlined  a theoretical framework
suitable for handling  relativistic finite nuclear system. The next step
would be to deal
with concrete applications: for example we can use our scheme  to
ascertain the antinucleonic components in the nuclear wave function,
which has never been properly
assessed. Furthermore, as already mentioned,
there is no reason preventing the extension of the
present formalism to describe a system made by quarks and gluons, because the
constrains we obtain on the self-energy only stem from Lorentz and
translational invariance. In fact the structure of a quark self-energy
in a nucleon should be simpler, because the excitation
spectrum of the nucleon is simpler than the one of a nucleus.

Clearly the deep inelastic scattering on both the nucleon and
the nucleus offer the best testing ground of the present formalism,
which might provide a help in achieving a better understanding of those
processes.

\appendix

\section{Appendix: Boosting the spectral amplitude}

We remind the definition
\begin{equation}
\phil=<(A-1)_l,{\bf k}-{\bf p}|\psi(0)|A,{\bf p}>\;.
\end{equation}
Let $\Lambda$ be a pure boost which transforms the state $|A,{\bf p}>$ to
a state at rest, i.e.
\begin{equation}
\Lambda|A,{\bf p}>=|A,{\bf p}=0>\quad.
\end{equation}
Obviously
\begin{equation}
\phil=<(A-1)_l,{\bf k}-{\bf p}|\Lambda^{-1}\left(\Lambda\psi(0)\Lambda^{-1}
\right)\Lambda|A,{\bf p}>\quad.
\end{equation}
We also know that
\begin{equation}
\Lambda\psi(0)\Lambda^{-1}=S\psi(0)\quad.
\end{equation}
where
\begin{equation}
S=e^{(-i/4)\sigma_{\alpha\beta}\omega^{\alpha\beta}}
\end{equation}
and
\begin{equation}
\sigma_{\alpha\beta}=\frac{i}{2}[\gamma_\alpha,\gamma_\beta]\quad.
\end{equation}
the parameters of a pure boost having the form
$\omega^{\alpha\beta}\to\omega^{0i}$.

The matrix $S$ transforming
a particle at rest (with mass $m$) into one
moving with momentum $\bf q$ is
\begin{equation}
S({\bf q})=\sqrt{\frac{E+m}{2m}}\left(\matrix{1&\frac{\bfm{\sigma}
\cdot{\bf q}}{E+m}\cr\frac{\bfm{\sigma}
\cdot{\bf q}}{E+m}&1}\right)
\end{equation}
with $E=\sqrt{q^2+m^2}$.
The velocity of the moving particle is
\begin{equation}
v=\frac{q}{\sqrt{q^2+m^2}}=\frac{q}{E}
\end{equation}
In our case the boost is characterized by the velocity
\begin{equation}
{\bf v}=\frac{{\bf p}}{\sqrt{p^2+M^2_A}}
\end{equation}
It is now immediate to conclude that
\begin{equation}
\Lambda\psi(0)\Lambda^{-1}=S(\frac{m}{M_A}{\bf p})\psi(0)
\end{equation}

Finally we need $\Lambda |(A-1)_l,{\bf p}-{\bf k}>$.
It clearly transforms in another state having the
same internal quantum numbers but a transformed momentum, namely
\begin{equation}
{\bf p}-{\bf k}\to {\bf q}=\frac{\sqrt{p^2+M^2_A}}{M_A}({\bf p}-{\bf k})
-\frac{\sqrt{({\bf p}-{\bf k})^2-(M_{A-1}^l)^2}}{M_A}{\bf p}
\end{equation}

We define in term of $\bf q$ the quantity
\begin{equation}
\tilde\phi({\bf q})=<(A-1)_l,{\bf q}|\psi(0)|A,0>
\end{equation}
and the previously derived transformation properties provide us the
factorization
\begin{equation}
\phil=S\left(\frac{m}{M_A}{\bf p}\right)\tilde\phi({\bf q})
\end{equation}

\vskip1.5cm
\centerline{\bf ACKNOWLEDGEMENTS}
\vskip0.5cm

One of us (G. V.) wishes to thank the University of Valencia for the
hospitality during the completion of this work.

\end{document}